 \documentclass[final,1p,times,twocolumn]{elsarticle}

\usepackage{amssymb}

\pdfoutput=1
\usepackage{color}
\newcounter{bla}

\journal{Computer Physics Communications}

\begin{document}

\begin{frontmatter}



\title{SALMON: Scalable Ab-initio Light-Matter simulator for Optics and Nanoscience}


\author[a]{Masashi~Noda}
\author[b]{Shunsuke~A.~Sato}
\author[c]{Yuta~Hirokawa}
\author[d]{Mitsuharu~Uemoto}
\author[a]{Takashi~Takeuchi}
\author[d]{Shunsuke~Yamada}
\author[d]{Atsushi~Yamada}
\author[e]{Yasushi~Shinohara}
\author[e]{Maiku~Yamaguchi}
\author[a]{Kenji~Iida}
\author[f]{Isabella~Floss}
\author[g]{Tomohito~Otobe}
\author[b]{Kyung-Min~Lee\footnote{Deceased 2 August, 2017.}}
\author[a]{Kazuya~Ishimura}
\author[d]{Taisuke~Boku}
\author[h]{George~F.~Bertsch}
\author[a]{Katsuyuki~Nobusada\footnote{Deceased 15 January, 2018.}}
\author[d]{Kazuhiro~Yabana$^*$}

\cortext[author] {Corresponding author.\\\textit{E-mail address:} yabana@nucl.ph.tsukuba.ac.jp}
\address[a]{Institute for Molecular Science, Okazaki, Japan}
\address[b]{Max Planck Institute for the Structure and Dynamics of Matter, Hamburg, Germany}
\address[c]{Graduate School of Systems and Information Engineering, University of Tsukuba, Japan}
\address[d]{Center for Computational Sciences, University of Tsukuba, Japan}
\address[e]{Graduate School of Engineering, The University of Tokyo, Tokyo, Japan}
\address[f]{Vienna University of Technology, Vienna, Austria}
\address[g]{National Institutes for Quantum and Radiological Science and Technology, Kyoto, Japan}
\address[h]{Institute for Nuclear Theory and Physics Department, University of Washington, Seattle, USA}

\begin{abstract}
SALMON (Scalable Ab-initio Light-Matter simulator for Optics and Nanoscience, http://salmon-tddft.jp) is a software package 
for the simulation of electron dynamics and optical properties of molecules, nanostructures, and crystalline solids based on 
first-principles time-dependent density functional theory. The core part of the software is the real-time, real-space calculation
of the electron dynamics induced in molecules and solids by an external electric field solving the time-dependent Kohn-Sham equation.
Using a weak instantaneous perturbing field, linear response properties such as polarizabilities and photoabsorptions in
isolated systems and dielectric functions in periodic systems are determined. Using an optical laser pulse, the ultrafast electronic 
response that may be highly nonlinear in the field strength is investigated in time domain. 
The propagation of the laser pulse in bulk solids and thin films can also be included in the simulation via coupling the electron dynamics in many microscopic
unit cells using Maxwell's equations describing the time evolution of the electromagnetic fields. The code is efficiently parallelized so that it may
describe the electron dynamics in large systems including up to a few thousand atoms. The present paper provides an 
overview of the capabilities of the software package showing several sample calculations.
\end{abstract}


\end{frontmatter}



{\bf PROGRAM SUMMARY}

\begin{small}
\noindent
{\em Program Title:} SALMON: Scalable Ab-initio Light-Matter simulator for Optics and Nanoscience     \\
{\em Licensing provisions(please choose one): Apache-2.0 }                                   \\
{\em Programming language:}   Fortran 2003                                \\

{\em Nature of problem (approx. 50-250 words):}\\
Electron dynamics in molecules, nanostructures, and crystalline solids induced by an external 
electric field is calculated based on first-principles time-dependent density functional theory.
Using a weak impulsive field, linear optical properties such as polarizabilities, photoabsorptions, 
and dielectric functions are extracted. Using an optical laser pulse, the ultrafast electronic response
that may be highly nonlinear with respect to the exciting field strength is described as well. 
The propagation of the laser pulse in bulk solids and thin films is considered by 
coupling the electron dynamics in many microscopic unit cells using Maxwell's equations describing 
the time evolution of the electromagnetic field.   \\

{\em Solution method (approx. 50-250 words):}\\
Electron dynamics is calculated by solving the time-dependent Kohn-Sham equation 
in real time and real space. For this, the electronic orbitals are discretized on a uniform Cartesian grid in three dimensions.
Norm-conserving pseudopotentials are used to account for the interactions between the valence electrons and the ionic cores.
Grid spacings in real space and time, typically 0.02 nm and 1 as respectively, determine the spatial and temporal resolutions of the simulation results.
In most calculations, the ground state is first calculated by solving the static
Kohn-Sham equation, in order to prepare the initial conditions.
The orbitals are evolved in time with an explicit integration algorithm such as a truncated
Taylor expansion of the evolution operator, together with a predictor-corrector
step when necessary.
For the propagation of the laser pulse in a bulk solid, Maxwell's equations 
are solved using a finite-difference scheme. By this, the electric field of the laser pulse and the 
electron dynamics in many microscopic unit cells 
of the crystalline solid are coupled in a multiscale framework.


\end{small}

\section{Introduction}

Optical responses to weak electromagnetic fields are characterized by linear susceptibilities 
such as polarizabilities for isolated systems and dielectric functions for extended systems. 
Accurate calculations of the susceptibilities have been one of the central issues in first-principles 
electronic structure calculations. Time-dependent density functional theory (TDDFT) 
\cite{Runge1984, Ullrich2012, Marques2012} combined with a linear response formalism is
known to provide descriptions of reasonable accuracy with moderate computational cost.
For molecules, linear-response TDDFT has been widely used to explore electronic excitations and
optical responses as one of the options in quantum chemistry calculations \cite{Casida2009, Laurent2013}. 
For solids, linear-response TDDFT provides reasonable descriptions for dielectric functions
if one employs advanced energy functionals such as a hybrid functional \cite{Paier2008}.
More elaborate first-principles approaches based on many-body perturbation theories such as the 
GW plus Bethe-Salpeter equation approach \cite{Rohlfing2000, Onida2002, Attaccalite2011, Deslippe2012} have also been successful.

There are many areas on the current frontier of optical science that require theory beyond
perturbative treatment of electromagnetic radiation with matter.
For example, time-domain measurements of ultrafast phenomena using pump-probe techniques 
are widely conducted and are recognized to be indispensable in current optical sciences.
The time resolution of the measurements has reached a few tens of attosecond, and time-domain
measurements of electron motion in molecules and solids have been extensively investigated  \cite{Krausz2009}. 
Light-matter interactions in nano-materials that do not allow dipole approximation have also attracted
much interest \cite{Lal:2007aa}. 
An important example is the optical near-field excitation of nanostructures \cite{Novotny:7JuyLhEr} 
that is a nonpropagating light localized at the nanostructures.
It is unique in a sense of its nonuniformity in space, and causes unusual optical responses such as electric 
field enhancement, making optically forbidden transitions possible, second-harmonic excitations \cite{Yamaguchi:2016hp}, 
and wavevector excitations.

As a simulator to meet the above demands, we have developed the computer code SALMON 
(Scalable Ab-initio Light-Matter simulator for Optics and Nanoscience) \cite{SALMON_web}.
It models the electron dynamics induced by a pulsed electric field at the first-principles level based on TDDFT.
SALMON calculates time evolutions of valence electron orbitals solving the time-dependent
Kohn-Sham (TDKS) equation using a real-time and real-space computational method 
\cite{Yabana1996, Yabana1999, Nobusada2004, Yabana2006, Nobusada2007}.
Norm-conserving pseudopotentials are used for the interactions between valence electrons and ions \cite{Troullier1991}. 
Expressing orbitals on the real space grid, SALMON describes electron dynamics in systems with various 
spatial structures: molecules and nano-particles using isolated and crystalline solids using periodic boundary conditions.

Electron dynamics calculations based on TDDFT started in the mid-1990s.
Some of the authors of the present paper have been involved in the development of the method 
from the beginning. A linear response calculation algorithm using an impulsive distortion was introduced
in \cite{Yabana1996}, inspired by a method developed in nuclear physics \cite{BLOCKI1979163, Bertsch1983}.
A corresponding algorithm for crystalline solids was developed in \cite{Bertsch2000}.

Over the last decade, two programs to solve the TDKS equation in real time and real space, GCEED and
ARTED, were produced in Japan. SALMON has been born merging these two programs.
GCEED (Grid-based Coupled Electron and Electromagnetic field Dynamics) \cite{Noda2014} 
has been developed at the Institute for Molecular Science as a program to describe electron dynamics 
in molecules and nanostructures. The code is efficiently parallelized by splitting the spatial area as well as 
the orbitals so that it is capable of calculating electron dynamics of large systems composed of a few thousand atoms. 
ARTED (Ab-initio Real Time Electron Dynamics simulator) \cite{Sato2014JASSE, Hirokawa2016, Hirokawa2018} has been developed mainly 
at the University of Tsukuba as a program to describe electron dynamics in crystalline solids. 
It has a unique function of describing the propagation of a pulsed light in a bulk solid combining the 
electron dynamics calculations in many unit cells of the crystalline solid with Maxwell's equations for the 
electromagnetic field of the pulsed light \cite{Yabana2012}. Merging GCEED and ARTED, 
we intend to develop SALMON as a software with wide capabilities in optical science, applicable to molecules,
nanostructures, and bulk materials in a unified way.

There are several other programs that describe electron dynamics based on TDDFT.
OCTOPUS \cite{Andrade2012} is a well known program package, utilizing the same numerical scheme 
as SALMON, real time and real-space grid representation.
In fact, both codes, OCTOPUS and SALMON, have a common root, a collaborative work around 2000 
\cite{Yabana1996, Bertsch2000}. 
Furthermore, there are implementations of real-time TDDFT in plane-wave codes. FPSID describes molecular dynamics 
as well as electron dynamics \cite{Sugino1999}. Recently, real-time TDDFT was also implemented in Qbox, 
and a large-scale calculation has been reported \cite{Draeger2017}.
Using atomic orbitals, real-time TDDFT has been implemented \cite{Takimoto2007, Meng2008, PEMMARAJU2018} 
in SIESTA \cite{SIESTA2002} that
utilizes numerical atomic orbitals, and in Elk FP-LAPW \cite{Elk_web}.
Real-time TDDFT has also been included in standard quantum chemistry codes such as NWChem \cite{Lopata2011}.
Among these implementations, we intend to develop SALMON as a software focusing on light-matter interactions
as primary areas of applications including high-intensity laser science, nano-optics, near-field excitations, etc.

The construction of the paper is as follows. In Sec.\ 2, we give an overview of the code
and explain how to build and run it. In Sec.\ 3 and 4, we present computational methods and
typical calculations of light-induced electron dynamics for isolated and periodic systems, respectively. 
In Sec.\ 5, the coupled calculation of electron dynamics and light field propagation is presented and
a summary is given in Sec. 6.

\section{Overview of SALMON}

\subsection{Capabilities of SALMON}

In SALMON, electron dynamics is described within a spatial area of a cuboid (a rectangular parallelepiped). 
Norm-conserving pseudopotentials are used to account for the interactions between the valence electrons and 
the ionic cores. The orbitals of the valence electrons are discretized on a uniform Cartesian grid 
inside the cuboid. At present, two boundary conditions are permitted: vacuum beyond all the cuboid faces;
or periodic in all directions.

Regarding the exchange-correlation potential and energy, the adiabatic approximation is applied:
the same functional form as that used in the ground state calculation is used for the energy and the potential,
replacing density, current, and kinetic energy density with those constructed from the time-dependent
orbitals. 

Since most time-dependent calculations start from the ground state solution, users usually first solve
the static Kohn-Sham equation to determine the initial state. 
Quantities obtained from the ground state calculation include orbital functions, discrete orbital energies for 
isolated systems and energy bands for extended periodic systems, density of states, and forces acting on the ions.

The electronic orbitals are evolved under an external, time-dependent electric field
that is assumed to be spatially uniform. Note that the field inside a medium also includes the polarization field of the medium itself.
There are different classes of external fields to comply with the specific interest of the user.
Impulsive weak fields serve to explore linear susceptibilities such as polarizability and photoabsorption 
in molecules, and dielectric functions in solids. The distinguishing feature of this real time calculation is 
the capability of obtaining the spectrum covering the whole spectral range from a single time-evolution calculation. 
Compared with matrix-diagonalization approaches commonly used in quantum chemistry codes, 
it is computationally advantageous especially for large systems because of the low size scaling and for metallic systems 
in which many electronic excited states contribute to the optical response. 
Employing an optical electric field, simulations of experiments using ultrashort laser pulses are feasible. 
The time profile of the exciting laser field is defined by input parameters specifying intensity, frequency, duration,
polarization direction, and carrier envelope phase of the pulse. To mimic pump-probe measurements, 
the field may be composed of two successive pulses.
If is also possible to provide a numerical table specifying the exciting field and allowing for all possible time profiles in the simulation. 
Quantities that can be obtained from time evolution calculations include electron density and
current as functions of space and time, polarization, electronic excitation energies, the number density of 
excited electron-hole pairs, and forces acting on the ions as functions of time.

To treat large systems, efficient parallelization schemes are implemented in the code. 
For isolated systems, parallelization can be applied on a spatial level splitting the cuboid as well as for the orbitals. 
Using a massively parallel supercomputer, one may calculate large systems composed of as many as 
a few thousand of atoms. For periodic systems, 
the orbital updating as well as calculations at different $k-$points in the Brillouin zone may be done in parallel.

SALMON has a unique feature of describing the propagation of a pulsed light field in bulk solids coupling Maxwell's 
equations with the TDKS equation in a multiscale implementation. Since it is necessary to carry out a large 
number of electron dynamics calculations simultaneously, this multiscale simulation requires extensively large
computational resources. At present, SALMON supports the propagation along one spatial direction such as a linearly polarized laser pulse incident 
normally on a bulk surface. In this case, only a one-dimensional chain of unit cells is needed which is currently feasible on modern supercomputers.

\subsection{How to build SALMON}

SALMON is mainly written in Fortran 2003 and supports both single and multiprocessing platforms using OpenMPI. 
In order to build SALMON, C and Fortran compilers (Fortran2003 or later) on the 
conventional UNIX environment, and a LAPACK (Linear Algebra PACKage) are required. For the LAPACK,
the build with intel MKL (Math Kernel Library) or Fujitsu SSL-II (Scientific Subroutine Library-2) has been thoroughly tested. 
Since SALMON utilizes CMake for the standard build method, CMake of version 3.0.2 or later is required.

To start the build procedure, create the temporary directory \verb|build_temp| at the top directory of the package
and move to the created directory:
\begin{verbatim}
mkdir build_temp
cd build_temp
\end{verbatim}
Then, enter the following two commands filling \verb|<ARCHITECTURE>| and \verb|<DIRECTORY>| appropriately:
\begin{verbatim}
python ../configure.py --arch=<ARCHITECTURE> --prefix=<DIRECTORY>
make
make install
\end{verbatim}
The configure script automatically sets the compiler flags and the path to the required libraries. 
The \verb|--arch| switch specifies the platform of your own environment. 
The entire list of the supported \verb|<ARCHITECTURE>| is provided on the SALMON website \cite{SALMON_web}.

By default, SALMON is built on a parallel platform. 
To build it on a single processing platform, add the flag \verb|--disable-mpi| to the \verb|configure.py|.
Other options of the configure script may be found by typing the command \verb|configure.py --help|.
If the build is successful, the binary executable file \verb|salmon.cpu| is created in the 
\verb|<DIRECTORY>/bin| directory of the package.

\subsection{How to run SALMON}

The binary executable requires an input file and files for the pseudopotentials to run.
SALMON utilizes Fortran namelist where the namelist variables should be specified in the input file. 
In Table \ref{table1}, the main namelist groups of SALMON are listed.
The complete list of the namelist variables is given in the file \verb|manual/input_variables.md|.
To prepare a new input file, it is convenient to start with one of the sample input files that are 
given in \verb|Exercises| on the SALMON website \cite{SALMON_web} and to modify it.

\begin{table}[tb]
\begin{center}
\caption{Namelist groups for SALMON inputs.}
\begin{tabular}{| l | l |} \hline
namelist group & purposes \\
\hline
\verb|&calculation|  &  Specify calculation modes. \\
\verb|&control|  &  Set options for restart. \\
\verb|&units| & Specify the unit system for input. \\
\verb|&parallel| & Set parameters related to parallelization. \\
\verb|&system| & Specify the material information. \\
\verb|&atomic_red_coor| & Specify atomic positions in reduced coordinate system.\\
\verb|&atomic_coor| & Specify atomic positions in ordinary Cartesian coordinate system. \\
\verb|&pseudo| & Specify information on pseudopotentials. \\
\verb|&functional| & Specify exchange-correlation functional. \\
\verb|&rgrid| & Specify real-space grid. \\
\verb|&kgrid| & Specify k-points for periodic system. \\
\verb|&tgrid| & Specify time grid for real-time propagation. \\
\verb|&propagation| & Choose the algorithm for the time propagation. \\
\verb|&scf| & Set parameters related to scf loop. \\
\verb|&emfield| & Set parameters of applied electric fields. \\
\verb|&multiscale| & Set parameters related to coupled Maxwell + TDDFT calculations. \\
\verb|&analysis| & Specify information and options for output files. \\
\verb|&hartree| & Specify the method to calculate Hartree potential. \\
\verb|&opt| & Set parameters related to structure optimization. \\
\verb|&md| & Set parameters related to molecular dynamics calculations. \\
\hline
\end{tabular}
\label{table1}
\end{center}
\end{table}

SALMON supports several formats for pseudopotential files 
including \verb|.cpi| (fhi98pp) \cite{Fuchs1999}.
In addition,  \verb|.fhi| format, which is a part of previous atomic data files for the
ABINIT code \cite{Gonze2016}.
Pseudopotential files for a few elements can be found in the \verb|Exercises|. 

Before the execution, the environmental variable, \verb|OMP_NUM_THREADS|, has to be set if thread parallelization is used. 
In the case of bash, it is set by 
\begin{verbatim}
export OMP_NUM_THREADS=(number of threads)
\end{verbatim}
In the case of csh or tcsh, it is set by 
\begin{verbatim}
setenv OMP_NUM_THREADS (number of threads)
\end{verbatim}

To run SALMON on a single node, type the following command:
\begin{verbatim}
$ salmon.cpu < inputfile.inp > fileout.out
\end{verbatim}
To run it on multiple nodes, the execution command is
\begin{verbatim}
$ mpiexec -n NPROC salmon.cpu < inputfile.inp > fileout.out
\end{verbatim}
where \verb|NPROC| is the number of MPI processes.
To run SALMON on many-core processors like Intel Xeon Phi, 
the execution command is
\begin{verbatim}
$ mpiexec.hydra -n NPROC salmon.mic < inputfile.inp > fileout.out
\end{verbatim}

\section{Calculations for isolated systems}

\subsection{TDKS equation}
SALMON computes the electron dynamics with either isolated or periodic boundary conditions.
We first consider a description for isolated systems. The TDKS equation including an external
electric field is given by \cite{Runge1984, Yabana1996, Nobusada2004, Nobusada2007} 
\begin{equation} \label{TDKS}
i\hbar \frac{\partial}{\partial t} \psi_p({\bf r}, t)  = \bigg[ -\frac{\hbar^2}{2m}\nabla^2 + V_{\rm ion} 
+ e^2\int \frac{\rho({\bf r'}, t)}{\left| {\bf r - r'} \right|} d{\bf r'}
+ V_{\rm xc} + V_{\rm ext} \bigg]\psi_p({\bf r}, t) , 
\end{equation}
where $m$ is the electron mass and $e$ is the elementary charge. 
$V_{\rm ion}$, $V_{\rm xc}$, and $V_{\rm ext}$ are electron-ion potential, exchange-correlation potential, 
and external potential of the applied electric field, respectively. 
At present, SALMON only treats spin-saturated systems for which each orbital is occupied by two electrons
and the electron density $\rho$ is given by 
\begin{equation}\label{rho}
	\rho(\mathbf{r},t) = 2 \sum_{p=1}^{N/2}|\psi_p(\mathbf{r},t)|^2 
\label{rho_psi}
\end{equation}
where $N$ is the number of valence electrons.
For the electron-ion potential $V_{\rm{ion}}(\mathbf{r})$, norm-conserving pseudopotentials \cite{Troullier1991} 
are used for each atomic component of the system with the Kleinman-Bylander 
separable approximation \cite{Kleinman1982}.
The exchange-correlation potential $V_{\rm xc}$ is treated in the adiabatic local-density approximation
(LDA) using the Perdew and Zunger \cite{Perdew1981} static potential
\begin{equation}
V_{\rm xc}[\rho](\mathbf{r},t) \approx V_{\rm xc}^{\rm LDA}[\rho(\mathbf{r},t)].
\end{equation}

We assume the dipole approximation for the external potential $V_{\rm ext}$,
\begin{equation}
V_{\rm ext}(\mathbf{r},t) = e \mathbf{E}(t) \cdot \mathbf{r}.
\end{equation}
For the external electric field $\mathbf{E}(t)$, two kinds of perturbing fields can be applied.
One is an instantaneous weak field that is used to investigate linear response properties of the system.
The other one is a pulsed electric field that is used to simulate interactions between an ultrashort 
laser pulse and the valence electrons in the system.

The impulsive field applied at $t=0$ is expressed as
\begin{equation}
{\mathbf E}(t) = F_0 \delta (t) \mathbf{e}_{\nu},
\end{equation}
where $F_0$ specifies the magnitude of the impulse of the field and
$\mathbf{e}_{\nu}$ is the unit vector in $\nu$ direction with $\nu \in \{x,y,z\}$ specifying the polarization direction.
Immediately after the field is applied, the orbitals change from the ground state orbitals $\phi_p(\mathbf{r})$ to 
$\psi_p(\mathbf{r},t=+0) = \exp (-ieF_0 r_{\nu}/\hbar) \phi_p(\mathbf{r})$.
The TDKS Eq.\ (\ref{TDKS}) is then solved using the distorted orbitals as initial condition.
The time-dependent induced dipole moment $d_{\nu}(t)$ is calculated as
\begin{equation}
d_{\mu}(t)= -e \int r_{\mu} \left\{ \rho(\mathbf{r},t)-\rho_0(\mathbf{r}) \right\} \mathrm{d}\mathbf{r}
\end{equation}
where $\rho_0(\mathbf{r})$ is the ground state density.
The distribution function of oscillator strength $\mathrm{d}f_{\mu\nu}(E)/\mathrm{d}E$, which is
related to the photoabsorption cross section of the system, is calculated as 
\begin{equation} \label{distribution}
\frac{\mathrm{d}f_{\mu\nu}(E)}{\mathrm{d}E} = \frac{2mE}{\pi\hbar^3}{\rm Im}\int_0^T f \left( \frac{t}{T} \right)d_{\mu}(t) e^{iEt/\hbar} \mathrm{d} t,
\end{equation}
where $T$ is the time duration of the calculation and $f(x)$ is a masking function that reduces
the undesirable effects caused by the abrupt termination of the calculation at time $T$.
We use the cubic polynomial, $f(x)=1-3x^2+2x^3$.

Equation (\ref{distribution}) produces a photoabsorption spectrum in a wide
energy range once we have performed the time-propagation of the orbitals.
This fact is one of the advantages compared to conventional quantum chemistry computation methods, 
which usually require matrix diagonalization of a large dimension to obtain the photoabsorption spectrum,
especially at higher energies. 

We next consider the case of irradiation by a pulsed light.
In SALMON, electric fields derived from the vector potential
\begin{equation} \label{pulse}
\mathbf{A}(t)
=
\mathrm{Im} \; \left\{
  \mathbf{A}_0
  \cos^2\ \left[
  \frac{\pi}{T} \left(
    t - \frac{T}{2}
  \right)
  \right]
  \exp\left[
    - i \omega \left(
      t - \frac{T}{2}
    \right)
    - 2 \pi i x_\mathrm{CEP}
  \right]
\right\}
\hspace{5mm}
\left( 0 < t < T \right),
\end{equation}
are available where $\mathbf{A}_0$ specifies the amplitude and the polarization direction of the field.
A general elliptically polarized pulse may be defined using complex components for $\mathbf{A}_0$.
$T$, $\omega$, and $x_{\mathrm CEP}$ specify the duration, average frequency, and carrier 
envelope phase of the pulsed field, respectively. One may use two successive pulses of time profiles given by Eq. (\ref{pulse}) 
with a time delay to carry out simulations that mimic pump-probe experiments. 
Alternatively to the pulsed field of Eq. (\ref{pulse}), it is possible to specify the electric field using a numerical table.

\subsection{Parallelization scheme}
For calculations of isolated systems, an efficient parallelization scheme is implemented in order to make calculations of large systems feasible. 
Details of this scheme have been described in \cite{Noda2014, Noda2015}. 

For the calculations of the Kohn-Sham orbitals $\psi_i(\mathbf{r},t)$, parallelizations based on divisions according to
both the orbital index $i$ and the spatial region of the coordinate $\mathbf{r}$ are implemented.
The most suitable parallelization scheme, however, differs depending on the stages of the calculation. 
In the ground state calculation, the parallelization by spatial divisions is more efficient than that by orbitals 
because calculations sequential in the orbitals appear in the Gram-Schmidt orthogonalization. 
For the time propagation calculations, on the contrary, the orbital parallelization is more efficient
since the propagation of each orbital can be performed independently.

During the time propagation, there are two kinds of communications in each time step.
One is necessary for the calculation of the electron density from the orbitals using Eq.\ (\ref{rho_psi}).
For this, \verb|MPI_allreduce| is used to add up all electron density contributions 
among nodes that treat orbitals within common spatial regions.
The other one is needed to exchange orbital information at 
border regions between the adjacent areas which is done via
\verb|MPI_isend| and \verb|MPI_irecv|.

In the decision of the node allocation for orbitals and spatial divisions,
it is important to minimize the communication costs and to avoid a load imbalance
at the nonlocal part of the pseudopotential. On the basis of our experience, 
we recommend to assign the nodes to $2 \times 2 \times 2$ spatial divisions 
and to use the remainder for the parallelization of the orbitals especially for large systems that have 
about 200 or more orbitals.

In our code, an automatic process assignment is implemented. At present, it works only if the total number of processes
may be factorized in prime factors of 2, 3, and 5 and is applied when the user does not specify the namelist \verb|parallel|.
It is also possible to set the assignment of nodes manually specifying the parallelizations of the orbitals 
and the grid. There are, however, some restrictions in the assignment: see the SALMON website \cite{SALMON_web} for details.

For the Hartree and the exchange-correlation potentials, calculations are distributed to 
all available nodes utilizing \verb|MPI_allgatherv|. 
OpenMP parallelization is implemented mainly for the grid parallelization. Users may use hybrid parallelization 
without any specifications in the input file.

\subsection{Examples}
\begin{figure}[tb]
\begin{center}
\includegraphics[width=13cm]{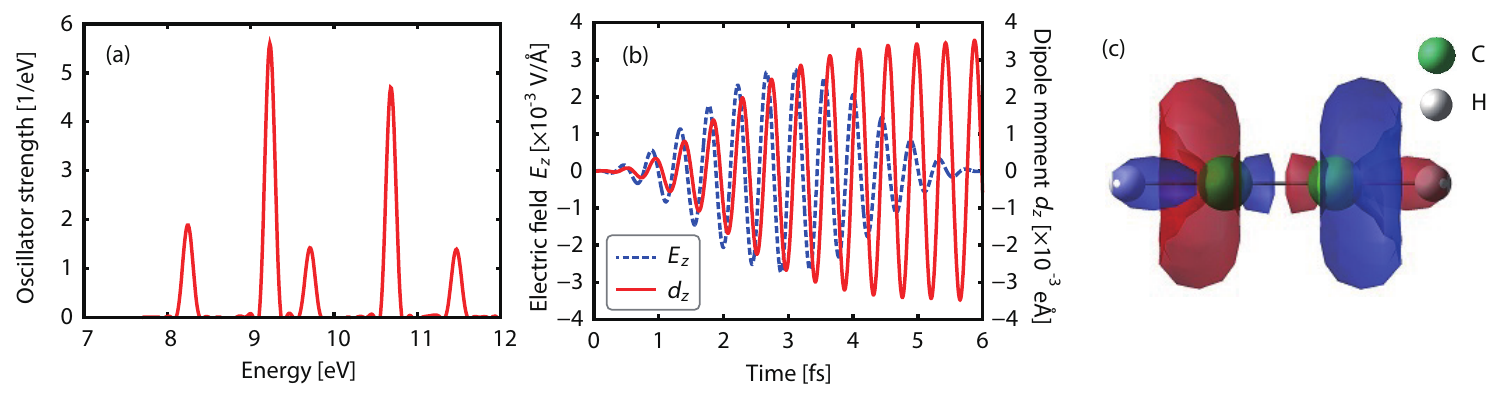}%
\caption{\label{C2H2-RT} (a) Photoabsorption spectrum of an acetylene molecule. (b) Applied electric field (blue-dashed) and 
the induced dipole moment (red-solid) in the acetylene molecule. (c) A snapshot of the induced charge density 
at 5.875 fs. The red and blue contours represent 1$\times$10$^{-4}$ \AA$^{-3}$ and -1$\times$10$^{-4}$ \AA$^{-3}$, respectively.}
\end{center}
\end{figure}

Let us first show the photoabsorption spectrum of an acetylene molecule as an example for linear response calculations of molecules. 
Figure \ref{C2H2-RT} (a) shows the calculated photoabsorption spectrum.
A strong resonance appears at the frequency of 9.23 eV$/\hbar$ in the spectrum. 
Next, we explore the response of the molecule to a few-cycle laser pulse. The pulse shape is shown
by the blue-dashed line in Fig.\ \ref{C2H2-RT}(b): we use a central frequency of 9.23 eV$/\hbar$, a 
pulse duration of about 6 fs, and a peak intensity of 10$^8$ W/cm$^2$. 
The time profile of the induced dipole moment is shown as red-solid line in Fig.\ \ref{C2H2-RT} (b). 
Note that the electronic excitation persists after the applied field ends, as to be expected
for an on-resonance excitation.
Users may obtain results for electron density, electron localization 
function, and local electric fields at each time step if the option is specified. 
As an example, the induced charge density directly after the laser pulse irradiation 
is shown in Fig.\ \ref{C2H2-RT} (c). 

We next present photoexcitation dynamics of Ag@Si core-shell nanostructures.
Core-shell nanostructures have great potential in materials science, since their optical response
can be tuned by changing the inner-core and outer-shell radii.
To understand the optical properties, photoexcitation electron dynamics simulations are useful, particularly 
to elucidate the light-matter interactions at the interface region between the core and the shell structures.
We should note, however, that those core-shell structures that are expected to be utilized in light responsive 
devices are generally large in size extending typically several nanometers or more.
Therefore, realistic first-principles simulations of core-shell structures are computationally highly demanding.

\begin{figure}[tb]
\begin{center}
\includegraphics[width=13cm]{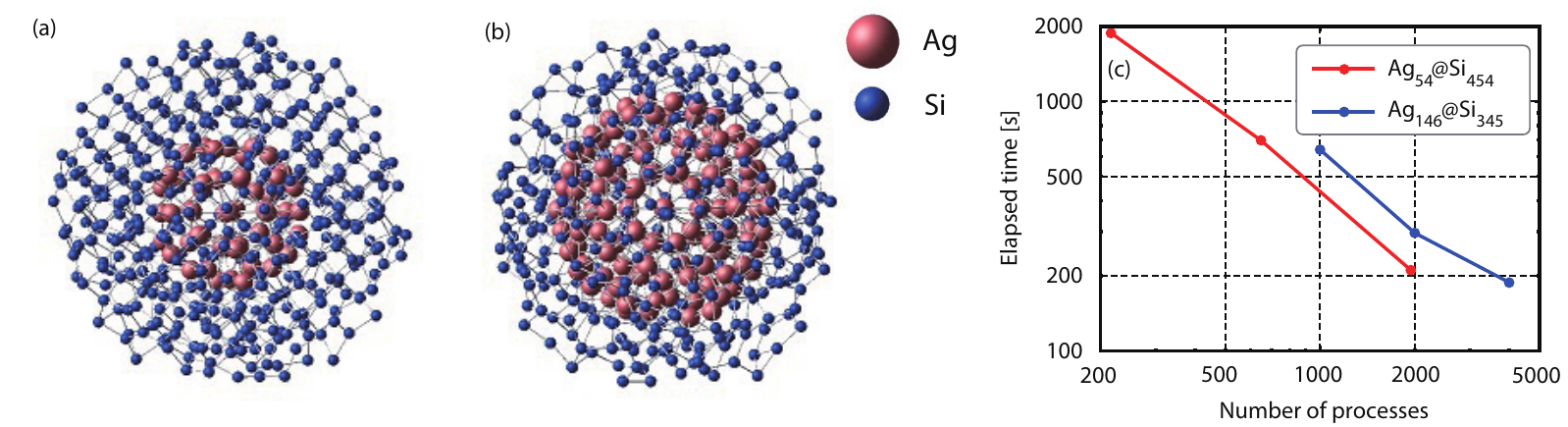}%
\caption{\label{coreshell} Geometrical structures of (a) Ag$_{54}$@Si$_{454}$ and (b) Ag$_{146}$@Si$_{345}$. 
(c) Elapsed times for 1000 time 
steps in the time evolution calculations of Ag$_{54}$@Si$_{454}$ and Ag$_{146}$@Si$_{345}$. }
\end{center}
\end{figure}

\begin{figure}[tb] 
\includegraphics[width=13cm]{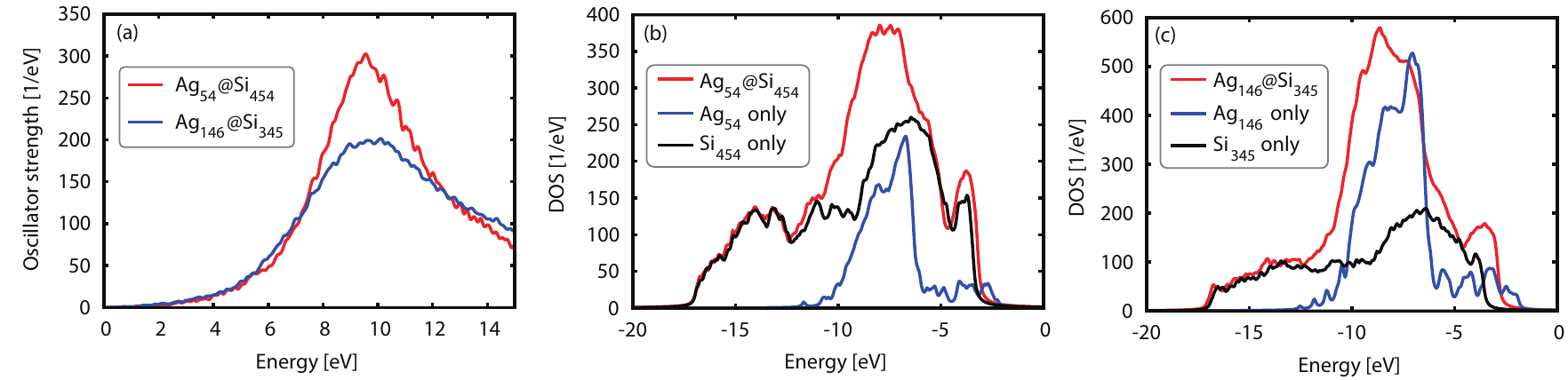}%
\caption{\label{cs-st-dos} (a) Oscillator strength distributions of Ag$_{54}$@Si$_{454}$ and Ag$_{146}$@Si$_{345}$. 
(b) Density of states of Ag$_{54}$@Si$_{454}$, isolated Ag$_{54}$, and isolated Si$_{454}$. 
(c) Density of states of Ag$_{146}$@Si$_{345}$, isolated Ag$_{146}$, and isolated Si$_{345}$.}
\end{figure}

We show calculations of linear optical properties of Ag$_{54}$@Si$_{454}$ and Ag$_{146}$@Si$_{345}$ 
core-shell structures whose geometric structures are shown in Fig.\ \ref{coreshell} (a) and (b), respectively. 
The total size of both core-shell structures considered are similar to each other having diameters of 25 \AA, 
but the ratios of inner-core to outer-shell volumes significantly differ: the diameter of the Ag core is 10 \AA\  for Ag$_{54}$@Si$_{454}$ 
and 15 \AA\  for Ag$_{146}$@Si$_{345}$. 
The total number of orbitals for Ag$_{54}$@Si$_{454}$ and Ag$_{146}$@Si$_{345}$ are 1,205 and 1,493, respectively. 
Calculated oscillator strength distributions of Ag$_{54}$@Si$_{454}$ and Ag$_{146}$@Si$_{345}$
are shown in Fig.\ \ref{cs-st-dos} (a). As can be seen from the figure, the peak of the photoabsorption shifts
towards higher energies as the Ag core becomes larger. Calculated densities of states are shown in Fig.\ \ref{cs-st-dos} (b) and (c),
comparing isolated Ag cores and Si shells with the interacting core-shell structures. 

These calculations are carried out on the K computer at the RIKEN Advanced Institute for Computational Sciences.
In K computer, each node consists of a SPARC64 VIIIfx processor (8 cores, 2.0 GHz) and a 16 GB memory. 
The parallelization efficiency can be extracted from the dependence of the computational time on the number
of MPI processes shown in Fig.\ \ref{coreshell} (c).
For Ag$_{54}$@Si$_{454}$, the parallel efficiency of 1,944 MPI processes based on 216 MPI processes is 99.2\% and 
the CPU efficiency for 1,944 MPI processes is 12.06\%. 
For Ag$_{146}$@Si$_{345}$, the parallel efficiency of 4,000 MPI processes based on 1,000 MPI processes is 85.3\% and 
the CPU efficiency for 4,000 MPI processes is 9.20\%. 
 
\section{Calculations for periodic systems}

\subsection{TDKS equation}

In this section we explain how to calculate the electron dynamics in a unit cell of a crystalline solid 
induced by a time-dependent, spatially-uniform electric field $\mathbf{E}(t)$. 
The velocity gauge in which the field is described by the vector potential 
$\mathbf{A}(t) = -c \int^t dt' \mathbf{E}(t')$ is adopted.
In this gauge, the lattice periodicity of the Kohn-Sham Hamiltonian is preserved at any time
so that the orbitals $\psi_p(\mathbf{r},t)$ may be expressed as
$\psi_p(\mathbf{r},t) = e^{i\mathbf{k}\mathbf{r}} w_{n\mathbf{k}}(\mathbf{r},t)$,
where $w_{n\mathbf{k}}(\mathbf{r},t)$ are Bloch orbitals having the periodicity
of the lattice. The orbital index $p$ separates into the band index $n$ 
and the crystalline momentum $\mathbf{k}$.

The time-dependent Bloch orbitals $w_{n\mathbf{k}}(\mathbf{r},t)$ satisfy the TDKS equation \cite{Bertsch2000},
\begin{equation} \label{TDKS_solid}
i\hbar \frac{\partial}{\partial t} w_{n\mathbf{k}}(\mathbf{r},t)
= \left\{ \frac{1}{2m} \left( -i\hbar \mathbf{\nabla} + \hbar \mathbf{k} + \frac{e}{c}\mathbf{A}(t) \right)^2
+ V_{\rm ion} + V_{\rm H} + V_{\rm xc} \right\} w_{n\mathbf{k}}(\mathbf{r},t),
\label{TDKScell}
\end{equation}
where $V_{\rm ion}$, $V_{\rm H}$, and $V_{\rm xc}$ are the norm-conserving pseudopotentials,
the Hartree, and the exchange-correlation potentials, respectively. All these potentials satisfy the
same lattice periodicity. We note that the nonlocal part of the pseudopotential depends on the 
vector potential to preserve the gauge invariance. For the exchange-correlation potential, the 
adiabatic approximation is applied. Besides a simple LDA, the 
Becke-Johnson (BJ) potential \cite{Becke2006}  and its variant by Tran and Blaha
\cite{Tran2009} are usable. They are known to reproduce the band gaps of various insulators 
reasonably well. 

We note that the field inside a bulk material is given by the sum of the external and the induced fields.
The induced field is caused by a surface polarization that depends on the shape of the bulk material. 
For most purposes, it will be appropriate to use the external field as the vector potential $\mathbf{A}(t)$
in Eq.\ (\ref{TDKScell}). We call this the transverse geometry. In our early publications, we presented 
several calculations in which the vector potential is given by the sum of the external and the
induced potentials in a geometry of a thin film \cite{Bertsch2000, Otobe2008} which we call the longitudinal geometry. 
As a means to calculate dielectric functions, it is confirmed that both choices give the same results \cite{Yabana2012}.

It has been argued that the exchange-correlation effects of infinite periodic systems cannot be treated
by periodic scalar potentials only but should be included in the vector potential as ${\mathbf A}_{xc}(t)$ \cite{Vignale1996}. 
However, the necessary extension to the vector potential has not yet been implemented in the code.

In the present code, a cuboid shape (rectangular parallelepiped) is assumed for the unit cell volume.
Therefore, appropriate supercells should be used for all systems with non-cuboid primitive unit cells.
The orbitals $w_{n\mathbf{k}}(\mathbf{r},t)$ are discretized on a uniform Cartesian grid, 
as it is done for isolated systems.
At present, the application is limited to nonmagnetic materials.

As in the case of isolated systems, two kinds of perturbing fields are prepared.
One is an impulsive weak field applied at $t=0$. In the velocity gauge adopted here,
it is realized by a shift of the vector potential,
\begin{equation} \label{impulsive_A}
\mathbf{A}(t) = \mathbf{e}_{\nu} A_0 \theta(t),
\end{equation}
where $A_0$ and $\mathbf{e}_{\nu}$ specify the magnitude and the direction of the distortion,
respectively, and $\theta(t)$ is the step function. The perturbing field induces
a macroscopic electric current density $\mathbf{I}(t)$ that is the electric current 
density averaged over the unit cell volume. It is given by,
\begin{equation} \label{macro_current}
\mathbf{I}(t) = \frac{-e}{\Omega} \int_{\Omega} d\mathbf{r}
\sum_{n\mathbf{k}} w^*_{n\mathbf{k}}(\mathbf{r},t)
\frac{1}{m} \left( -i\hbar\nabla + \hbar \mathbf{k} + \frac{e}{c}\mathbf{A}(t) \right)
w_{n\mathbf{k}}(\mathbf{r},t) + \mathbf{I}_{\rm ion\_nonlocal},
\end{equation}
where $\mathbf{I}_{\rm ion\_nonlocal}$ indicates the electric current that originates
from the nonlocal part of the pseudopotentials. The conductivity of the system is calculated by
\begin{equation}
\sigma_{\mu\nu}(\omega) = 
-\frac{c}{A_0}
\int^T_0 dt e^{i\omega t} f \left( \frac{t}{T} \right) I_{\mu}(t),
\label{sigma_ft}
\end{equation}
where $T$ is the duration of the time evolution, and $f(x)$ is a mask function smoothly going from 1 to 0.
The dielectric function is given by
$\epsilon_{\mu\nu}(\omega) = \delta_{\mu\nu} + 4 \pi i \sigma_{\mu\nu}(\omega)/\omega$.

We next consider the case of applying a pulsed electric field.
The same pulsed fields as explained in Eq.\ (\ref{pulse}) for isolated systems are usable,
including two-pulse irradiation to mimic a pump-probe experiment 
and arbitrary fields defined by a numerical table.

It should be noted that there are two methods to calculate the electronic excitation 
energy per unit cell volume, $E_{\rm ex}(t)$ \cite{Sato2015JCP}. One is to calculate directly the difference between 
the electronic energy at time $t$ and the ground state energy. For the other one the work
done by the applied electric field is determined via $W(t) = - \int^t dt' \, \mathbf{I}(t') \cdot \mathbf{E}(t')$.
In principle, the two methods are equivalent if the energy functional is available.
Only the latter method is implemented for exchange-correlation potentials that have not
been derived from known energy functionals.

\subsection{Examples}

\begin{figure}[tb]
\begin{center}
\includegraphics[width=10cm]{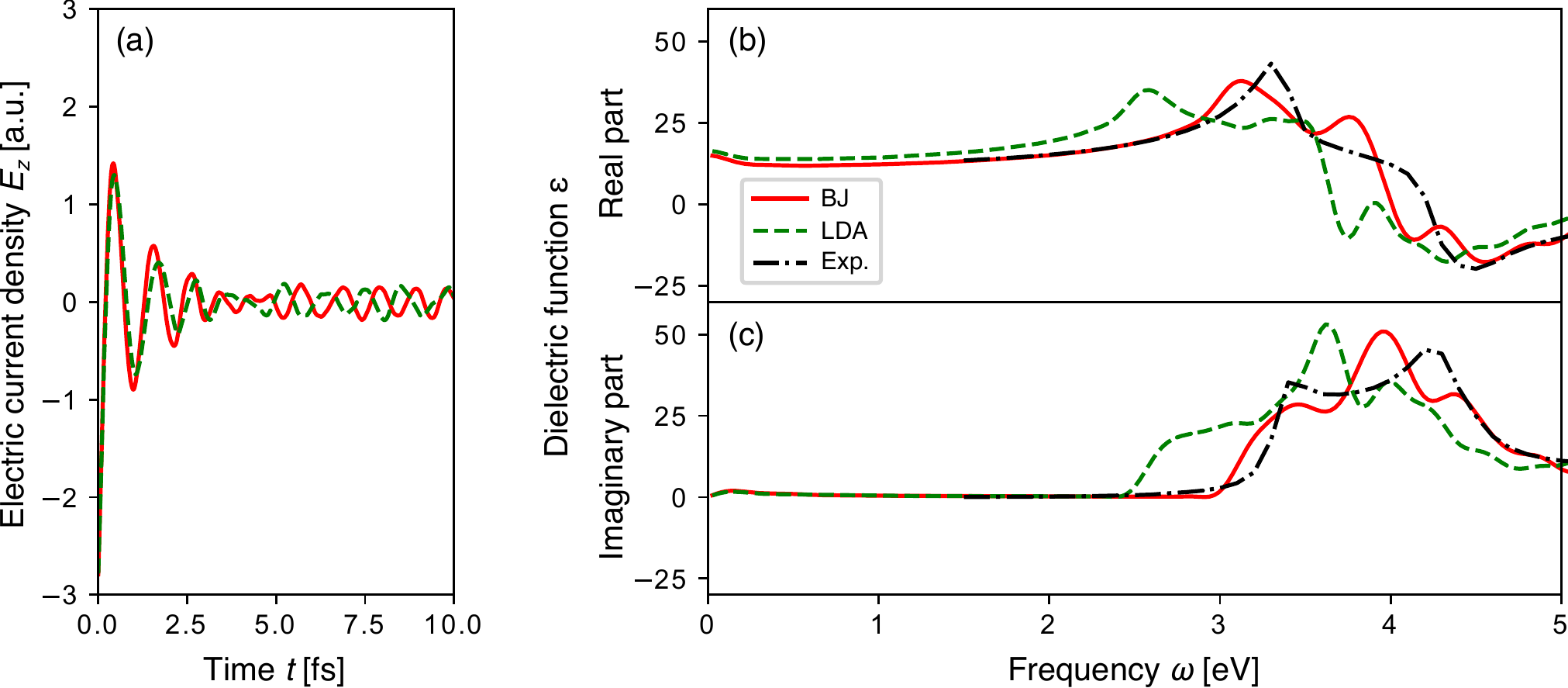}
\caption{\label{Si_dielectric} Linear response calculation in crystalline silicon. (a) Electric current induced by an impulsive field. 
(b) Real and imaginary parts of the calculated dielectric function 
are compared with measured values \cite{Aspnes1983}. 
}
\end{center}
\end{figure}

In Fig.\ \ref{Si_dielectric}, we show the calculation of the dielectric function of bulk silicon.
At time $t=0$, an impulsive field is applied leading to post-oscillations in the current density (panel (a)). 
>From this, the dielectric function (panels (b) and (c)) is determined as explained in the previous section.
In this calculation, a small constant current arising from tiny numerical inaccuracies is subtracted in Eq. (\ref{sigma_ft}),
$I_{\mu}(t) \rightarrow I_{\mu}(t) - {\bar I_{\mu}}$, to enforce the relation $\sigma_{\mu\nu}(0)=0$
that should be satisfied exactly for materials of finite bandgap.
Results using LDA (green-dashed) and BJ (red-solid)  potentials are compared with 
the measured values (black-dotted). As seen from the figure, the optical bandgap of silicon is 
reasonably reproduced using the BJ potential.

To show the electron dynamics induced by a 
few-cycle electric field we consider bulk silicon as target material (Fig.\ \ref{Si_pulse}). 
The average frequency of the pulse is set to 1.55 eV$/\hbar$,
the pulse duration in FWHM to 7 fs, and the maximum intensity of the field to 2.7 V/nm. 
We apply the LDA which gives 2.4 eV for the direct bandgap of silicon. 
Thus, the absorption of at least two photons is required for the electronic excitations.
Panel (a) shows the time profile of the applied electric field (black-dashed) and the induced 
electric current density (red-solid). Panel (b) shows the electronic excitation energy (blue-solid) 
and the number density of excited electron-hole pairs (green-dashed).
During the irradiation, the excitation energy and the number density of electron-hole pairs
increase gradually, and stay unchanged after the pulsed field ends.

In the calculations for crystalline solids, the number of orbitals to be evolved
in time is given by the number of orbitals in the unit cell times the number of $k-$points.
The convergence with respect to the number of $k-$points becomes slow when a longer pulse is used. 
The computational cost can be reduced by reducing the number of $k-$points accounting for the symmetry of the system,
but this has not yet been implemented in SALMON.
The present Si calculation with $16^3$ spatial grid points, $8^3$ $k-$points, and $10^4$ time steps costs 
3 CPU-hours for LDA and 16 CPU-hours for BJ calculations, using a single node of Knight-Landing processor 
with 68 cores \cite{Hirokawa2018}.

\begin{figure}[tb] \label{Si_pulse}
\begin{center}
\includegraphics[width=12cm]{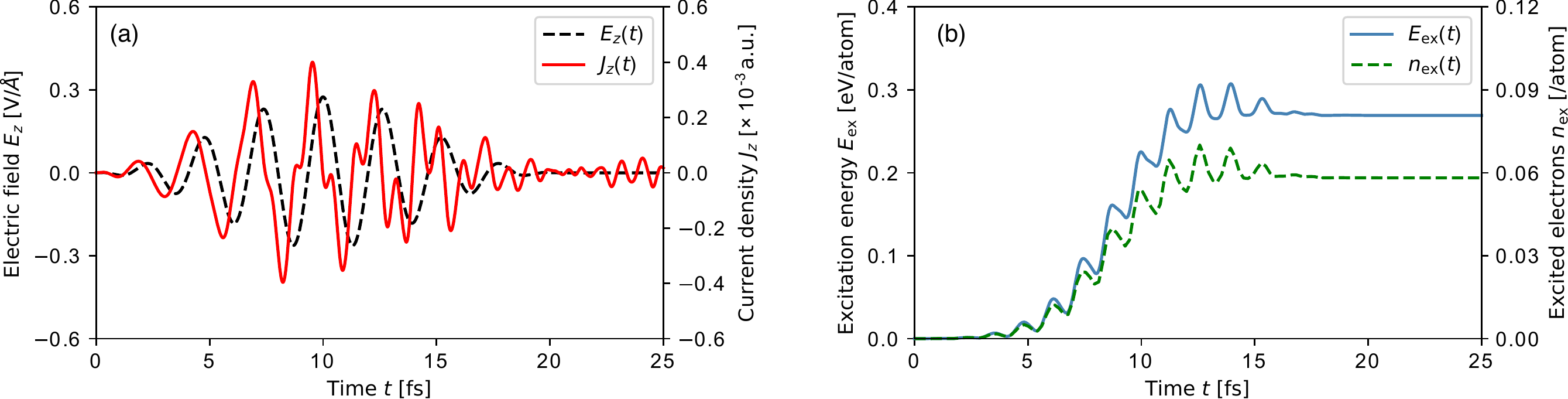}
\caption{\label{Si_pulse} Electron dynamics calculation in a unit cell of crystalline silicon. (a) Time profile of applied electric field (black-dashed)
and induced electric current density (red-solid). (b) Electronic excitation energy per unit volume (blue-solid) and the number density of electron-hole pairs (green-dashed).}
\end{center}
\end{figure}

\begin{figure}[tb]
\begin{center}
\includegraphics[width=10cm]{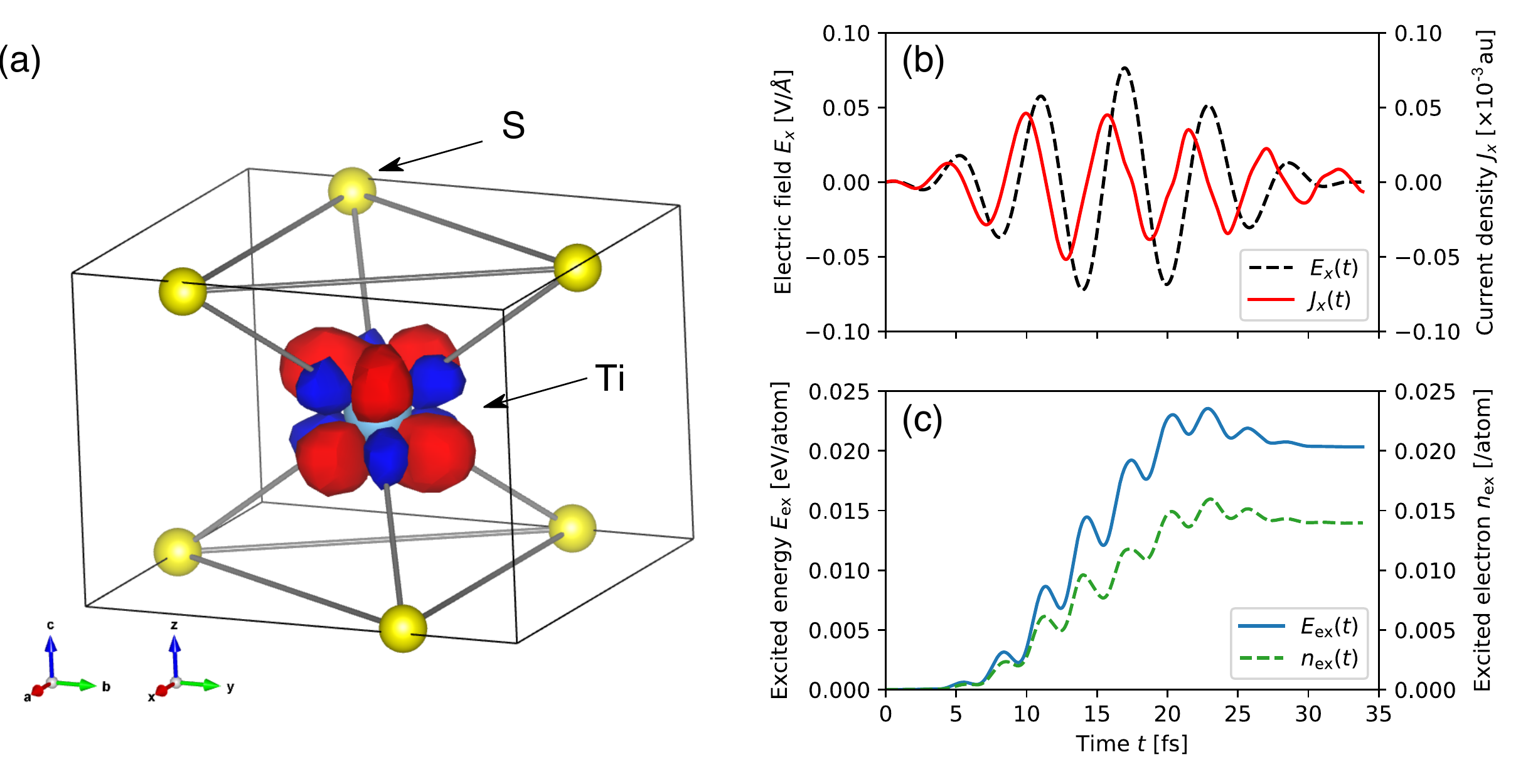}
\caption{\label{TiS2_pulse} Electron dynamics calculation in a unit cell of crystalline TiS$_2$. (a) 
Atomic positions and the nonlinear change of the electron density due to the laser irradiation are shown.
(b) Time profile of applied electric field (black-dashed ) and induced electric current density (red-solid)
are shown. (c) Electronic excitation energy (blue-solid) and the number density of excited electron-hole
pairs are shown as functions of time.}
\end{center}
\end{figure}

As another example, we show the electronic excitations in TiS$_2$ due to irradiation with an intense few-cycle pulse (Fig.\ \ref{TiS2_pulse}).
In this calculation, $32 \times 28 \times 28$ spatial grid points, $16 \times 10 \times 10$ $k-$points,
and 35,000 time steps are adopted.
TiS$_2$ has a layered structure with each Ti atom surrounded by six sulfide ligands in an
octahedral structure, as shown in Fig.\ \ref{TiS2_pulse} (a). The polarization direction of the 
applied electric field is chosen along the $c$-axis, the average frequency is 0.67 eV, and 
the pulse length is 12 fs in FWHM.
Panel (b) shows the applied electric field (black-dashed) and the induced current density (red-solid).
Panel (c) shows the electronic excitation energy (blue-solid) and the
number of excited electron-hole pairs (green-dashed) as functions of time. As seen from the ratio of the two latter
quantities, two-photon absorption is dominant in the excitation process.
In panel (a), the change of the electron density due to the pulse irradiation is shown at a time after the laser pulse ends.
It shows the nonlinear 
component obtained by subtracting a component that is linear in the field. As seen from the figure, the nonlinear
density change is located only around the Ti atom. 
It indicates electrons in the bond regions connecting 
Ti and S atoms move to t2g orbitals of Ti atoms.

Calculations using pulsed electric fields are beneficial to explore various ultrafast optical phenomena
in solids in time domain. This includes coherent phonon generation in crystalline solids \cite{Shinohara2010}, 
extremely nonlinear generation of ultrafast currents in dielectrics \cite{Wachter2014},
ultrafast changes of dielectric properties of solids \cite{Schultze2014, Otobe2016}, 
dielectric breakdown by optical pulses \cite{Otobe2008}, 
high harmonic generation in solids \cite{Otobe2009}, and many more.

\section{Calculations of pulsed-light propagation}

SALMON can be used to calculate the propagation of a light pulse in a bulk medium,
by simultaneously solving Maxwell's equations for the electromagnetic field and the
TDKS equation for the electron dynamics in individual unit cells of the medium
\cite{Yabana2012}. 
For this, the polarization induced by the microscopic electron dynamics is incorporated in the Maxwell
equations, while the magnetization, if present, is neglected. Since large computational resources are required, 
simulations of this mode are only feasible employing massively parallel supercomputers.
Details of the numerical implementation and the parallelization are described in \cite{Sato2014JASSE}.

In the following we explain the basic numerical method for the propagation of 
a linearly-polarized pulsed light incident normally on a flat surface of a bulk material.
The propagation axis perpendicular to the flat surface is chosen to be along the $x$ axis, the polarization direction
along the $z$ axis, and the surface of the bulk material is located at $x=0$. 
The $z$-component of the vector potential $A(X,t)$ that describes the electromagnetic fields of the pulsed light
satisfies the wave equation
\begin{equation} \label{Maxwell_ms}
\frac{1}{c^2} \frac{\partial^2}{\partial t^2} A(X,t) - \frac{\partial^2}{\partial X^2} A(X,t)
= \frac{4\pi}{c} I(X,t).
\end{equation}
Here $I(X,t)$ is the induced current density, assumed to be parallel to the 
polarization direction of the exciting field. This assumption holds for all materials showing inversion symmetry.
The wave equation (\ref{Maxwell_ms}) is solved on a uniform grid in the coordinate $X$.
At each grid point of $X$, its microscopic electron dynamics under the electric field given by 
$E(X,t)=-(1/c)\partial A(X,t)/\partial t$ is calculated. Therefore, its own set of Kohn-Sham orbitals $w_{n\mathbf{k},X}(\mathbf{r},t)$
is evolved in time using the TDKS equation
\begin{equation} \label{TDKS_ms}
i\hbar \frac{\partial}{\partial t}w_{n\mathbf{k},X}(\mathbf{r},t)
= \left\{ \frac{1}{2m} \left( -i\hbar\mathbf{\nabla} + \hbar \mathbf{k} + \frac{e}{c} \mathbf{e}_z A(X,t) \right)^2
+ V_{\rm ion, \it X} + V_{{\rm H},X} + V_{{\rm xc},X} \right\}
w_{n\mathbf{k},X}(\mathbf{r},t).
\end{equation}
Two spatial coordinates appear in the above equation, $X$ for the description of the vector potential, 
and $\mathbf{r}$ for the microscopic electron dynamics. 
We treat $X$ as a parameter for the microscopic description (\ref{TDKS_ms}) assuming $A$ to be uniform within the unit cell.
The electric current density $I(X,t)$ on the right hand side of Eq.\ (\ref{Maxwell_ms})
is calculated from the orbitals $w_{n\mathbf{k},X}(\mathbf{r},t)$ individually for each $X$.

At the beginning of the time propagation, the vector potential $A(X,t=0)$ is defined such 
that the laser pulse is located in the vacuum region in front of and approaching the surface. 
The electronic orbitals at each $X$ inside the material, $w_{n\mathbf{k},X}(\mathbf{r},t=0)$, 
are set to the ground state solution. Then, Eqs. (\ref{Maxwell_ms}) and (\ref{TDKS_ms}) are solved simultaneously.
The total energy in this simulation is composed of the energies of the electromagnetic fields and 
of the electronic excitations. It is useful to monitor the conservation of the energy to assess 
the accuracy of the results.

\begin{figure}[tb]
\begin{center}
\includegraphics[width=12cm]{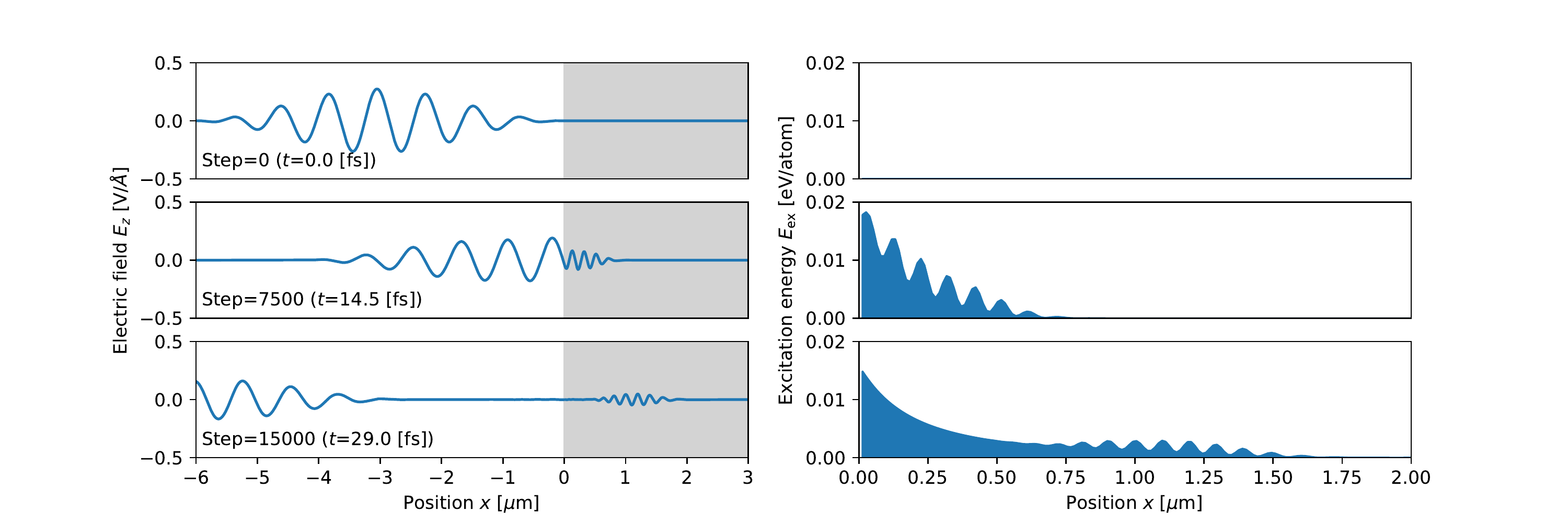}
\caption{\label{Si_MS} Coupled calculation of Maxwell's equations for pulsed light
and TDKS equation for electron dynamics in bulk silicon.  }
\end{center}
\end{figure}

As an example, we show in Fig.\ \ref{Si_MS} the irradiation of bulk silicon by a linearly polarized 
few-cycle laser pulse incident normally on the surface of the target. 
The left panels show the electric field at three times: before (top), during (middle) and after (bottom) the pulse hits the surface. 
As expected, the incident pulse is separated into transmitted and 
reflected parts. The frequency, pulse duration, and maximum intensity of the pulsed light are 
1.55 eV$/\hbar$,  20 fs, and $1.0 \times 10^{12}$ W/cm$^2$, respectively. LDA is used. 
The right panels show the electronic excitation energy per atom inside the medium 
at the times corresponding to the left panels. In the middle and bottom panels, two kinds of energy 
distribution are seen. One shows oscillations accompanying the pulsed field that indicates oscillatory
electronic motion caused by the pulsed field. The other comes from the irreversible energy transfer from the pulsed light to the electrons 
and decreases exponentially with increasing distance from the front surface.

In this calculation, 256 grid points are used for the coordinate $X$ to sample a silicon layer of about 3 $\mu$m thickness. 
At each grid point, the electron dynamics for a unit cell of silicon is calculated using the same
settings as described in the previous section. The overall computational cost can be estimated by the cost required for the electron dynamics simulation 
for a single unit cell times the number of grid points for $X$ inside the material.

This code capability will be useful to simulate experiments such as an intense, few-cycle laser pulse 
irradiating on a bulk dielectric or passing through a thin film. Applications have been made to 
analyze the energy transfer from the laser pulse to the electrons in transparent dielectrics 
that will be important to understand the initial stage of laser processing 
\cite{Lee2014, Sato2015, Sommer2016}, ultrafast change of dielectric properties of 
dielectrics \cite{Lucchini2016}, and generation of high harmonics in the propagation through dielectrics \cite{Floss2018}.

\section{Summary}

The present paper aims to provide an overview of SALMON
that is a scientific computer program to describe electron dynamics 
in molecules, nanostructures, and solids induced by external electric fields
based on first-principles time-dependent density functional theory.
The code is useful for a variety of problems relevant to light-matter interactions.
In the linear regime, calculations of polarizability and photoabsorption in molecules 
and nanostructures, and dielectric functions in solids are feasible.
In the nonlinear regime, ultrafast electronic excitations and electron dynamics induced by 
few-cycle laser pulses can be studied in detail. 

We intend to develop SALMON in two directions: 
One is to provide a numerical experiment platform to simulate forefront 
optical experiments precisely. This helps in understanding the underlying mechanism
of experiments, elucidating electron dynamics at the nano-meter spatial and
attosecond time scales. It also helps to find novel phenomena for which
direct measurements are difficult due, for example, to the lack of appropriate 
light sources.
The other is to provide an easy and ubiquitous access to a computational 
platform also for those who are not very familiar with first-principles calculations 
and, for instance, need some quantitative estimates before building complex experimental setups.

\section*{Acknowledgments}
This research was supported by JST-CREST under grant number JP-MJCR16N5,
and by MEXT as a priority issue theme 7 to be tackled by using Post-K Computer, 
and by JSPS KAKENHI Grant Numbers 15H03674, 26-1511, and 16K00175.
S.A.S. gratefully acknowledges fellowships by the Alexander von Humboldt Foundation.
I.F. was supported by the FWF Austria (SFB-041 ViCoM, SFB-049 NextLite and doctoral college W1243) and the IMPRS-APS.
Calculations are carried out at Oakforest-PACS at JCAHPC.

\label{}





\bibliographystyle{elsarticle-num}
\bibliography{salmon}







\end{document}